\documentclass[longauth]{aa} 

\usepackage{graphicx}
\usepackage{txfonts}
\usepackage{natbib}
\bibpunct{(}{)}{;}{a}{}{,} 
\usepackage{color}
\usepackage{newtxtext}
\usepackage[colorlinks=true,linkcolor=black,citecolor=blue,urlcolor=blue,draft=false]{hyperref}

\usepackage{todonotes}

\begin{document}

\title{Disk Evolution Study Through Imaging of Nearby Young
Stars (DESTINYS): A close low mass companion to ET Cha\thanks{Based on data obtained in ESO programs 1104.C-0415(E) and 70.C-0286(A)}} 

\titlerunning{DESTINYS - A close companion to ET Cha}

\author{C.~Ginski\inst{1,2}
\and F.~M\'enard\inst{3}
\and Ch. Rab\inst{4}
\and E.~E.~Mamajek\inst{5,6}
\and R.~G. van Holstein\inst{2,7}
\and M. Benisty\inst{8,9}
\and C.~F. Manara\inst{10}
\and R. Asensio Torres\inst{11}
\and A. Bohn\inst{2}
\and T. Birnstiel\inst{12,13}
\and P. Delorme\inst{9}
\and S. Facchini\inst{10}
\and A. Garufi\inst{14}
\and R. Gratton\inst{15}
\and M. Hogerheijde\inst{2,1}
\and J. Huang\inst{16}
\and M. Kenworthy\inst{2}
\and M. Langlois\inst{17}
\and P. Pinilla\inst{11}
\and C. Pinte\inst{18, 9}
\and \'{A}. Ribas\inst{7}
\and G. Rosotti\inst{2}
\and T.~O.~B. Schmidt\inst{19}
\and M. van den Ancker\inst{10}
\and Z. Wahhaj\inst{7}
\and L.~B.~F.~M. Waters\inst{20}
\and J. Williams\inst{21}
\and A. Zurlo\inst{22}
}

\institute{Anton Pannekoek Institute for Astronomy, University of Amsterdam, Science Park 904, 1098XH Amsterdam, The Netherlands \email{c.ginski@uva.nl}
\and Leiden Observatory, Leiden University, PO Box 9513, 2300 RA Leiden, The Netherlands
\and Univ. Grenoble Alpes, CNRS, IPAG, 38000 Grenoble, France
\and Kapteyn Astronomical Institute, University of Groningen, P.O. Box 800, 9700 AV Groningen, The Netherlands
\and Jet Propulsion Laboratory, California Institute of Technology, M/S 321-100, 4800 Oak Grove Drive, Pasadena, CA 91109, USA
\and
Department of Physics \& Astronomy, University of Rochester, 500 Wilson Blvd., Rochester, NY 14627, USA
\and
European Southern Observatory (ESO), Alonso de C\'ordova 3107, Vitacura, Casilla 19001, Santiago de Chile, Chile
\and 
Unidad Mixta Internacional Franco-Chilena de Astronom\'ia, CNRS, UMI 3386 and Departamento de Astronom\'ia, Universidad de Chile, Camino El Observatorio 1515, Las Condes, Santiago, Chile
\and
Univ. Grenoble Alpes, CNRS, IPAG, 38000 Grenoble, France
\and
European Southern Observatory, Karl-Schwarzschild-Strasse 2, 85748 Garching bei München, Germany
\and
Max Planck Institute for Astronomy, Königstuhl 17, 69117 Heidelberg, Germany
\and
University Observatory, Faculty of Physics, Ludwig-Maximilians-Universität München, Scheinerstr. 1, D-81679 Munich, Germany
\and
Exzellenzcluster ORIGINS, Boltzmannstr. 2, D-85748 Garching, Germany
\and
INAF, Osservatorio Astrofisico di Arcetri, Largo Enrico Fermi 5, I-50125 Firenze, Italy
\and
INAF-Osservatorio Astronomico di Padova,Vicolodell’Osservatorio 5, 35122 Padova, Italy
\and
Harvard-Smithsonian Center for Astrophysics, Cambridge, MA 02138, USA
\and
CRAL, UMR 5574, CNRS, Université de Lyon, École Normale Supérieure de Lyon, 46 Allée d'Italie, 69364, Lyon Cedex 07, France
\and
School of Physics and Astronomy, Monash University, Clayton, Vic 3800, Australia
\and
Hamburger Sternwarte, Gojenbergsweg 112, D-21029 Hamburg, Germany
\and
SRON Netherlands Institute for Space Research, Sorbonnelaan 2, 3584 CA Utrecht, The Netherlands
\and
Institute for Astronomy, University of Hawai'i at Mānoa, Honolulu, HI 96822, USA
\and
Facultad de Ingeniería y Ciencias,  Universidad Diego Portales, Av. Ejercito 441, Santiago, Chile
}

\date{Received:   ; accepted: }


\abstract
   { To understand the formation of planetary systems, one needs to understand the initial conditions of planet formation, i.e. the young gas-rich planet forming disks. Spatially resolved high-contrast observations are of particular interest, since substructures in disks, linked to planet formation, can be detected and close companions or even planets in formation embedded in the disk can be revealed.}
   {In this study we present the first result of the DESTINYS survey (Disk Evolution Study Through Imaging of Nearby Young Stars). DESTINYS is an ESO/SPHERE large program that aims at studying disk evolution in scattered light, mainly focusing on a sample of low-mass stars ($<$1Msun) in nearby ($\sim$200 pc) star-forming regions. In this particular study we present the observations of the ET\,Cha  (RECX 15) system, a nearby 'old' classical T Tauri star (5-8\,Myr, $\sim$100 pc), which is still strongly accreting. }
   {We use SPHERE/IRDIS in H-band polarimetric imaging mode to obtain high spatial resolution and high contrast images of the ET\,Cha system to search for scattered light from the circumstellar disk as well as thermal emission from close companions. We additionally employ VLT/NACO total intensity archival data of the system taken in 2003.}
   {We report here the discovery of a low-mass (sub)stellar companion with SPHERE/IRDIS to the $\eta$ Cha cluster member ET\,Cha. We are estimating the mass of this new companion based on photometry. Depending on the system age it is a 5\,Myr, 50\,$M_{Jup}$ brown dwarf or an 8\,Myr, 0.10\,$M_\odot$ M-type pre-main-sequence star. We explore possible orbital solutions and discuss the recent dynamic history of the system.}
   {Independent of the precise companion mass we find that the presence of the companion likely explains the small size of the disk around ET\,Cha. The small separation of the binary pair indicates that the disk around the primary component is likely clearing from the outside in, explaining the high accretion rate of the system.}


\maketitle

\section{Introduction}
 
 Gas giant planets are formed when the circumstellar disks around young stars are still rich in gas and dust. Dust in these disks must go through a very intense and rapid phase of growth, to transform ISM-like particles, sub-micron in size, to large bodies thousands of kilometer across. Irrespective of the exact details by which this happens, the formation of planets is intimately intertwined with the evolution of disks (see \citealt{Morbidelli2016} for a recent review).\\
 The results of surveys to measure the bulk properties and evolution timescales of disks indicate that disks around T Tauri stars dissipate on a typical timescale of 3~Myr (e.g., \citealt{2001ApJ...553L.153H, Hernandez2007, Fedele2010}). They also indicate that the mass available in solids (as estimated from mm-continuum observations) is at best of a few $M_\mathrm{jup}$ by 1-2\,Myr (\citealt{Ansdell2016, Pascucci2016}). Assuming a typical gas-to-dust mass ratio of 100, the typical total disk mass is of order 0.5\,\% of the central star mass (\citealt{2013ApJ...771..129A}).  These results, short timescales and limited amount of mass, place stringent constraints on the planet formation mechanisms (\citealt{Greaves2010,Najita2014,2018A&A...618L...3M}).\\
 New instruments providing high angular resolution and high contrast offer a new window to resolve the disks and study directly the presence and interaction of forming planets with their parental disks. However, at least for now, the results from these surveys are mostly relevant either for the brighter end of the young star sample when adaptive optics is used or to the largest and most massive disks when mm-interferometry is used (see e.g. \citealt{2018ApJ...869L..41A} and \citealt{2018A&A...620A..94G}).
 In this paper we report the first results of the Disk Evolution Study Through Imaging of Nearby Young Stars (DESTINYS), a large program carried out with SPHERE (\citealt{Beuzit2019}) at the ESO/VLT. DESTINYS will obtain deep, high contrast, polarized intensity images of a sample of 85 T Tauri stars in all nearby star forming regions to expand the current results towards the fainter members of the young stellar population.
In this study we present early observational results of the ET\,Cha system, located in the $\eta$ Chamaeleontis cluster.\\
The $\eta$ Chamaeleontis cluster is a nearby (d$\sim$97 pc), compact (core extent $\sim$1pc) and coeval (age $\leq$ 10 Myr) cluster of young stars \citep{Mamajek1999, Lawson2001, Herczeg2015}.
It contains approximately $\sim$20 low-mass members, a few of which have been confirmed by spectroscopy to sustain significant accretion \citep{Lyo2003,Murphy2011,Rugel2018}.
The most striking case is ET\,Cha (=ECHA~J0843.3-7905, RECX~15), which was the first low-mass member discovered through a photometric survey of the $\eta$ Cha cluster by  \citet{Lawson2002} \citep[the original low-mass members were all discovered via X-ray emission;][]{Mamajek1999}. 
ET\,Cha exhibited remarkably strong H$\alpha$ emission (EW(H$\alpha$) = -110\AA, \citealt{Lawson2002}) and strong 
far-IR 60$\mu$m and 100$\mu$m excess (\citealt{2009ApJ...701.1188S,Woitke2011}, IR counterpart is IRAS F08450-7854) - both indicative of an accreting classical T Tauri star. ET\,Cha stands out in the $\eta$ Cha cluster as the system with the most massive disk (M$_{dust}$=3.5$\times$10$^{-8}$\,M$_\odot$, \citealt{Woitke2019}) and the highest accretion rate \citep{Lyo2003}. The high accretion rate was confirmed by \cite{Lawson2004}, who measured $\sim$10$^{-9}$ M$_\odot$yr$^{-1}$ from H$\alpha$ equivalent width and \cite{Rugel2018} who found accretion rates between 5.8$\times10^{-10}$ M$_\odot$yr$^{-1}$ and 7.6$\times10^{-10}$ M$_\odot$yr$^{-1}$ from H$\alpha$, H$\beta$ and UV excess measurements. 
Interestingly, the disk was also estimated to be unusually compact by \citet{Woitke2011} who used global radiation thermo-chemical modelling. In particular, matching the low line flux of the [OI]63$\mu$m line and the non-detection of the CO 3-2 emission by APEX requires an outer disk radius of only $R_\mathrm{out} \lesssim 10\,\mathrm{au}$ \citep{Woitke2011}. 
This result was confirmed by more sensitive and spectrally resolved ALMA observations of $^{12}\mathrm{CO}\,J=3-2$, where the broad line width is consistent only with a disk outer radius of $5-10\,\mathrm{au}$ \citep{Woitke2019}. The continuum emission at $850\,\mathrm{\mu m}$, detected with ALMA, is consistent with a small and truncated but gas rich (gas-to-dust mass ratio of $\approx 3500$) circumstellar disk.\\
ET~Cha is one of the rare\footnote{We note that, while still rare, there is an increasing number of "old" systems with signs of ongoing accretion discovered in the recent literature. See \cite{Lee2020} for an example.} cases of a T Tauri star retaining its primordial gas-rich disk to a late age and as such it is an extremely interesting laboratory to study disk evolution. It is of particular interest why a rather old disk, which should have undergone viscous spreading, is seemingly so small in radial extent and why it still harbors a large amount of gas.\\ 
The paper is organized as follows. In section~2, we discuss the observation and data reduction. We analyze the data in section~3 and discuss the age of the ET\,Cha system in section~4. Related to this we discuss the presence of planetary mass companions in section~5. In section~6 we investigate the orbital architecture of the system given our previous findings. We finally discuss our new observations in the context of previous studies on the system in section~7 and conclude in section~8.

\section{Observations and Data Reduction}

\begin{table*}
 \centering
 \caption{Observing setup and average observing conditions SPHERE/IRDIS and archival NACO observation epochs.}
  \begin{tabular}{@{}cccccccc@{}}
  \hline 
 Epoch & Instrument & Coronagraph & Filter & DIT [s] & \# of frames & Seeing [\arcsec{}] & $\langle\tau_0\rangle$ [ms] \\
 \hline
21-01-2003	& NACO & no & H & 0.345 & 1971 & 0.55 & 5.5 \\
23-12-2019	& SPHERE & yes & BB\_H & 64 & 56 & 0.43 & 8.6 \\
23-12-2019	& SPHERE & no & BB\_H & 0.84 & 10 & 0.36 & 10.1 \\
\hline\end{tabular}
\label{tab: observing_setup}
\end{table*}

\subsection{SPHERE IRDIS Observations}

We observed ET\,Cha on 23rd of December 2019 with SPHERE/IRDIS in dual polarization imaging mode with pupil stabilization (\citealt{Langlois2014, deBoer2020, vanHolstein2020}).
The main observing sequence was conducted with the primary star behind a coronagraph with inner working angle of 92.5\,mas (\citealt{2011ExA....30...39C}) in the H-band. Individual integration times for this sequence were 64\,s per frame amounting to a total integration time of 59.7\, min. The main science sequence was preceded and followed by flux calibration frames taken with the primary star moved away from the coronagraph. Here shorter integration times of 0.84\,s per frame were set in order to prevent saturation. Total integration time for the flux reference frames amounts to 8.4\,s. Observational setup and weather conditions are summarized in table~\ref{tab: observing_setup}.\\
The data was reduced using the IRDAP pipeline (IRDIS Data reduction for Accurate Polarimetry, \citealt{vanHolstein2020}). The data reduction process is described in detail in \cite{vanHolstein2020}.  

\subsection{Archival VLT/NACO data}

ET\,Cha was observed with VLT/NACO (\citealt{2003SPIE.4841..944L, 2003SPIE.4839..140R}) in the H-band on 21st of January 2003.
Observing conditions were excellent with low seeing (0.55\arcsec{}) and above average atmosphere coherence time. The observations were conducted in field stabilized mode in the H-band with short exposure times of 0.345\,s. The total integration time amounted to 11.3\,min. The observing setup and conditions are summarized in table~\ref{tab: observing_setup}. The NACO "autojitter" template was used to move the star to different detector positions in order to enable an accurate sky background subtraction.\\
The data was reduced using the ESO eclipse software package and the jitter routine (\citealt{1999ASPC..172..333D}). Data reduction steps included sky subtraction, aligning of individual frames with a cross correlation routine and stacking. 

\section{Results}
\begin{figure}
\center
\includegraphics[width=0.48\textwidth]{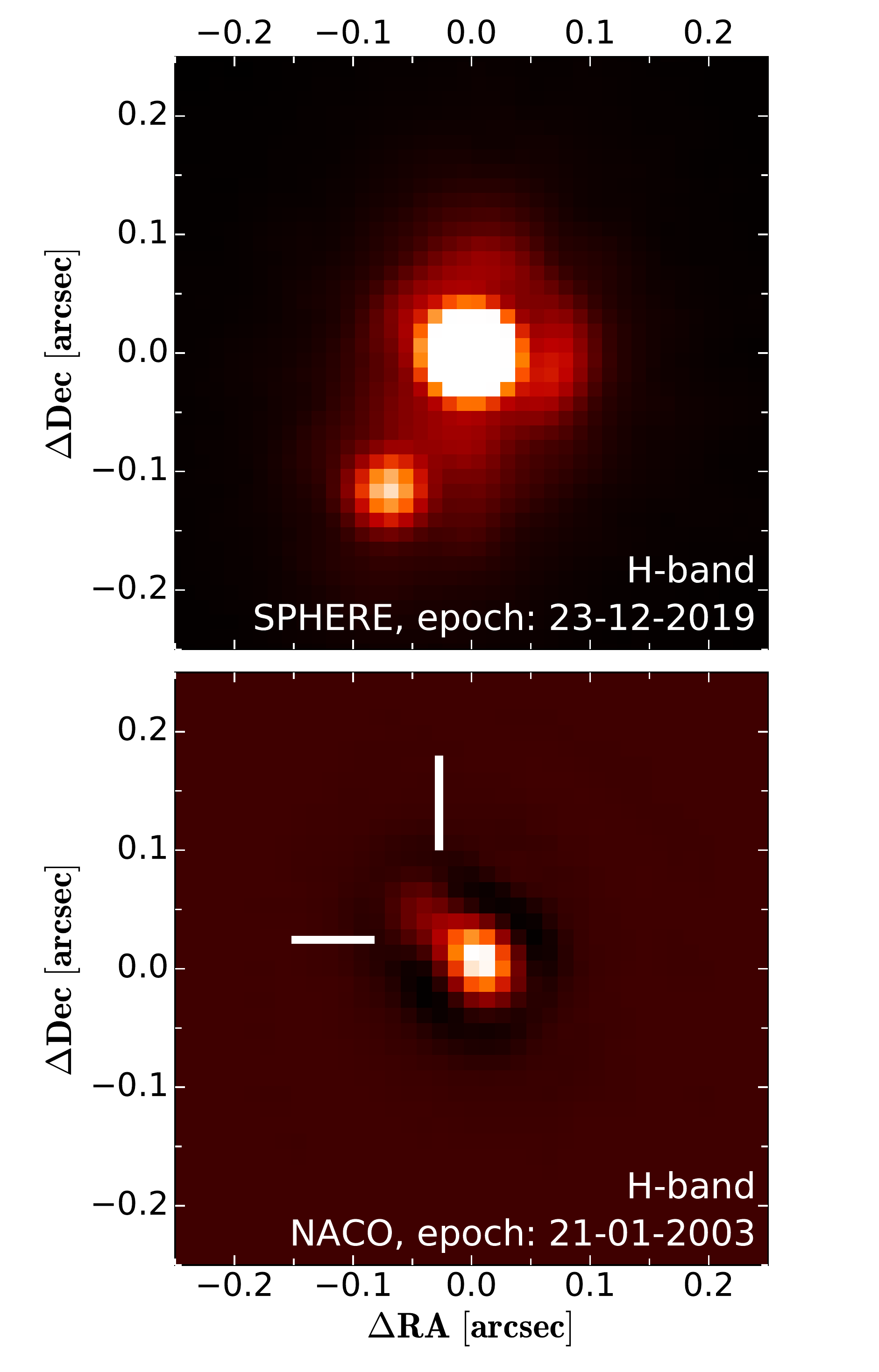} 
\caption{SPHERE/IRDIS and NACO observations of the ET Cha system. The companion is well resolved in the 2019 SPHERE epoch. Note that we show the primary star on a slightly saturated color scale in order to highlight the companion (the data is not saturated). In the 2003 NACO epoch the companion is close to the resolution limit of the instrument and shows as a strong asymmetrical extension to the primary star PSF. We have performed a high-pass filter to make the companion more clearly visible. We mark the companion position in the NACO image by two white bars. 
} 
\label{fig:sphere_images}
\end{figure}

\begin{figure*}
\center
\includegraphics[width=0.88\textwidth]{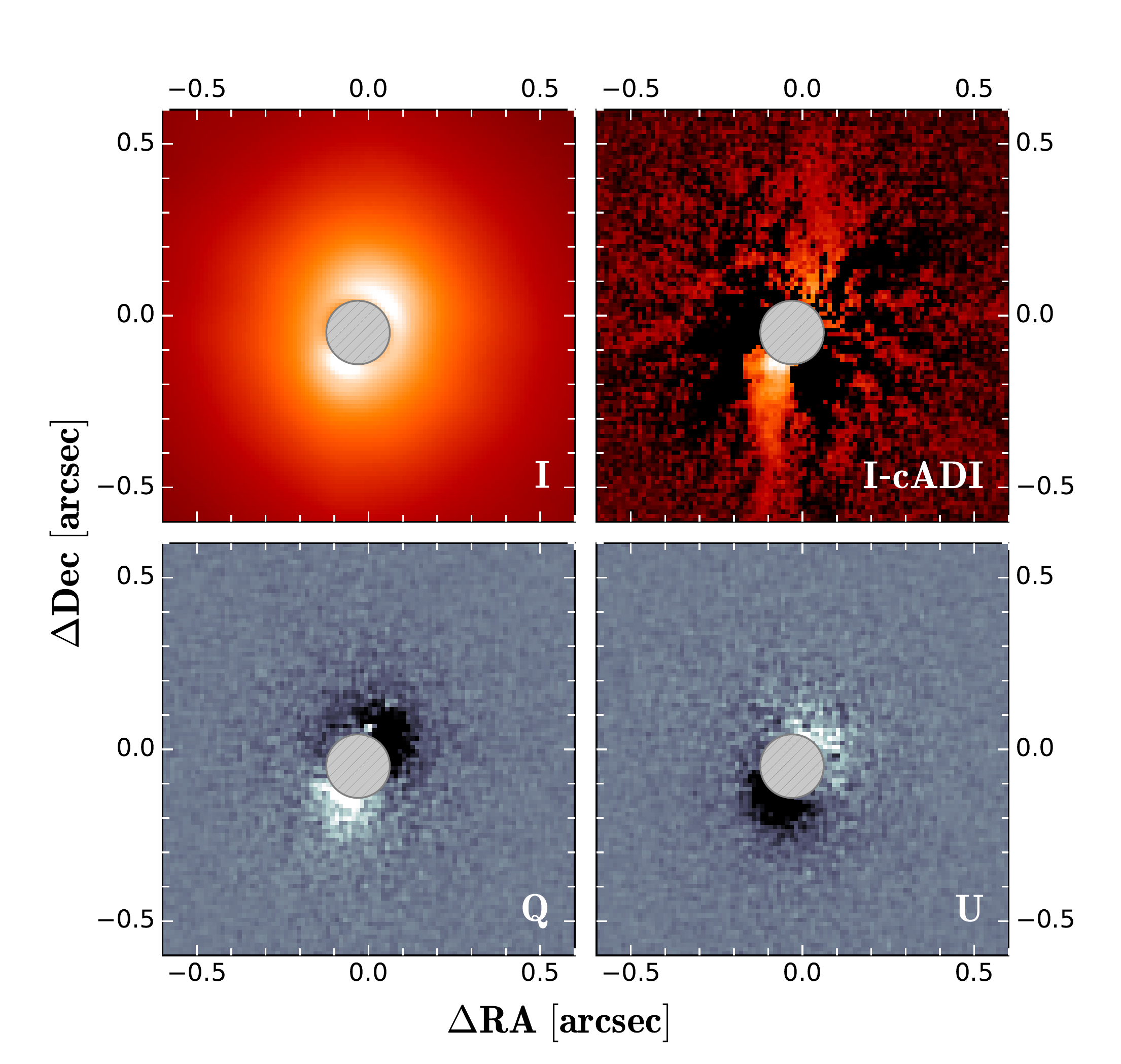} 
\caption{Coronagraphic images of ET\,Cha taken with SPHERE/IRDIS in our program. \textit{Upper left:} Stacked total intensity image. \textit{Upper right:} Total intensity image after classical angular differential imaging reduction. \textit{Bottom:} Stokes Q and U polarized flux images after polarization differential imaging.
The size of the coronagraphic mask is indicated with the grey, hashed circle. The positive-negative signal pattern is caused by unresolved stellar polarization of the primary star and/or the companion and not by a resolved circumstellar disk.
} 
\label{fig:coro}
\end{figure*}

In our SPHERE observations we find a close companion candidate to ET\,Cha. This companion candidate is most clearly visible in the flux reference frames taken without the coronagraph and shown in figure~\ref{fig:sphere_images}. The companion is very clearly detected in total intensity (i.e., polarized and unpolarized light combined), roughly 130\,mas South-East of the primary star. In the coronagraphic data the companion is detected, but with much lower S/N. This is because the primary star was not well centered behind the coronagraph, but the mask was rather placed roughly on the photocenter location between both sources and thus the companion candidate was inside of the inner working angle of the mask, i.e. it is suppressed by more than 50\,\%. 
We show the coronagraphic images after angular differential imaging was applied to subtract the primary star PSF in figure~\ref{fig:coro}. \\
We performed polarimetric differential imaging on the coronagraphic data to search for polarized scattered light from the circumstellar disk around ET\,Cha. We show the final Stokes Q and U images in figure~\ref{fig:coro}.
We find a positive-negative signal pattern along the direction in which stellar primary and companion are located. This is residual unresolved stellar polarization and not resolved signal from a circumstellar disk. For a disk we would expect a "butterfly" pattern associated with azimuthal polarization (see e.g. \citealt{Ginski2016}). This is not present here. Stellar polarization is discussed in detail in \cite{vanHolstein2020}. The changing sign that we observe between the residual signal from the companion (in the South-East) and the stellar primary (in the North-West) suggests that both of these sources show different absolute linear polarization. Since both sources are co-located at the same distance this can not be introduced by different column densities of interstellar dust. Instead it is likely that this is introduced by circumstellar material around the primary and/or companion. We find the most likely explanation that the known circumstellar disk around the primary star is inclined and thus introduces a break in symmetry in the unresolved system. This will naturally result in a residual polarization of the light that we receive. However since we do not know the exact geometry of the disk around the primary star we can only speculate on the degree of linear polarization that is introduced. We can thus not rule out that the light received from the companion is also intrinsically polarized, perhaps also by circumstellar material.\\
 We conclude that we did not detect any significant signal from a resolved circumstellar disk outside of the inner working angle of the coronagraph. Due to the mis-centering we conservatively estimate the inner working angle to be larger than the mask diameter, i.e. roughly 150\,mas. We note that this inner working angle is asymmetric with closer separations sampled in the North-West than in the South-East.\\
In addition to the new SPHERE observations we analyzed archival VLT/NACO data. In this data set, taken under excellent observing conditions, we find that the PSF of ET\,Cha is very clearly asymmetrically elongated towards the North-East (see figure~\ref{fig:sphere_images}, bottom panel). While such elongations are possible for other stars in the field-of-view due to the limited isoplanatic angle, they are not typical for the on-axis adaptive optics guide star itself. Furthermore elongations due to isoplanatic angle effects are typically point symmetric while this is clearly not the case here. We thus conclude that in this NACO data set we recover the companion candidate detected by SPHERE.
If this is the case, then the companion candidate has moved significantly relative to the primary star within the $\sim$17 year epoch difference between both data sets. In the following sections we extract the astrometry and photometry of the companion from the data and discuss the nature of the object.

\subsection{Astrometric analysis}

\begin{figure}
\center
\includegraphics[width=0.48\textwidth]{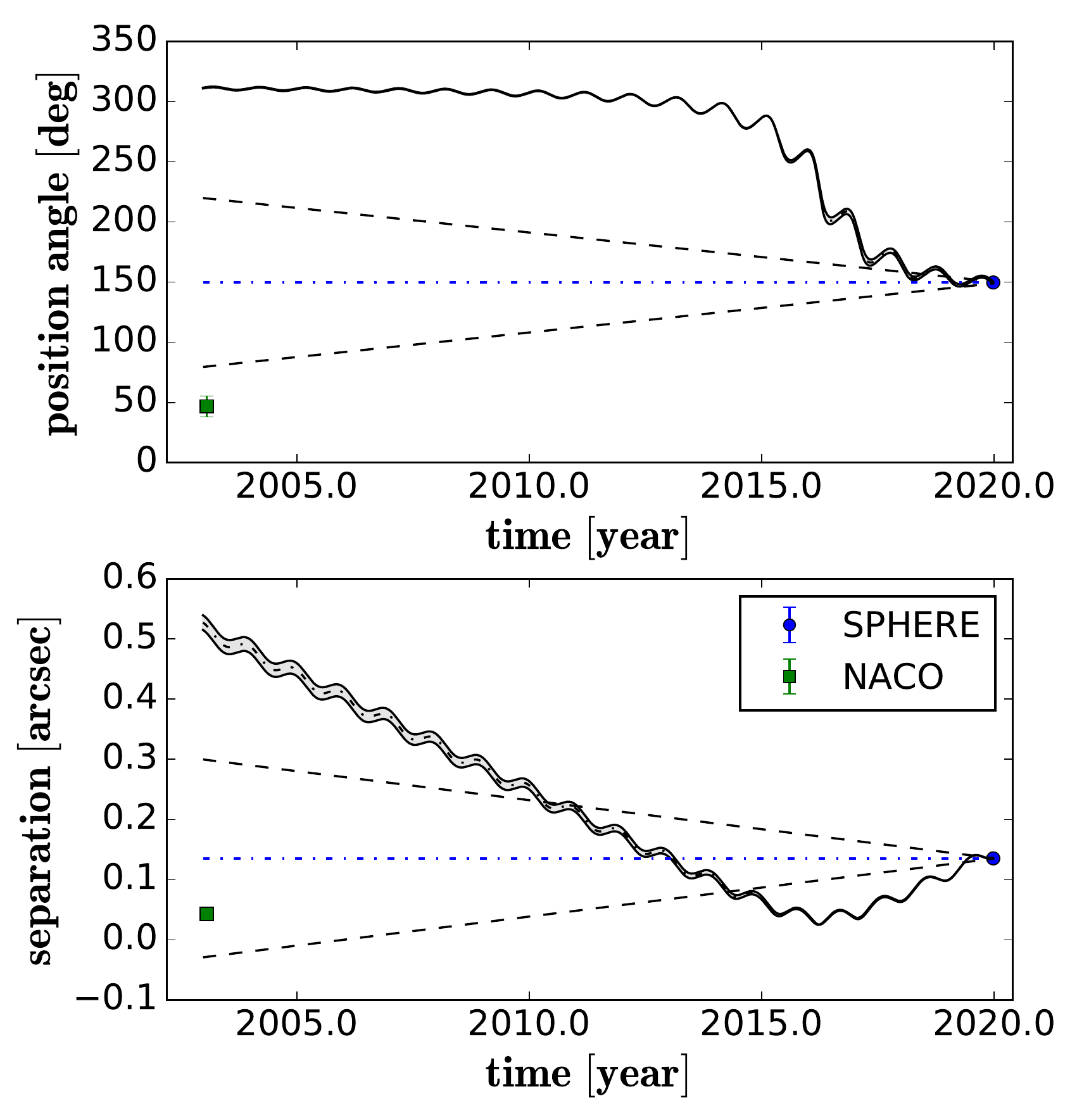} 
\caption{SPHERE/IRDIS and NACO astrometry of the detected companion relative to the primary star versus time.
Position angle is measured from North over East. The grey ribbon shows the expected location of a non-moving background object, while the dashed lines indicate possible circular orbital motion assuming a face-on orbit for position angle and an edge-on orbit for separation independently (these are mutual exclusive orbits to illustrate the maximum expected change in separation and position angle for the circular case).  } 
\label{fig:pm_analysis}
\end{figure}

Since the companion candidate was only well detected in the SPHERE flux reference frames without a coronagraph, we used only these for astrometric extraction.
The companion candidate is close to the primary star, which shows slightly asymmetric diffraction patterns, likely due to low-wind effect (ground wind speed was below 1\,ms$^{-1}$, see \citealt{2019Msngr.176...25C}).
It is thus difficult to remove either stellar PSF in absence of an independent reference PSF for the data set in order to measure individual stellar positions.
We therefore fitted both, the companion and the primary star position simultaneously.
As model we utilized two elliptical Moffat functions\footnote{In an upcoming publication (Ginski et al., in prep.), we extensively tested the influence of different fitted model functions on the retrieved astrometry for tight binary stars with SPHERE/ZIMPOL. We found that as long as the model has a well defined peak the astrometric result was virtually identical.}. We allowed for ellipticity in the Moffat in order to better fit small asymmetries in the stellar PSFs of companion candidate and primary star. Initial guesses of the position were assigned by eye and then a least-squares fitting approach was utilized as implemented in the \textit{astropy} model fitting package. The astrometric calibration for IRDIS was taken from \cite{2016SPIE.9908E..34M}\footnote{We note that the calibration was performed in standard imaging mode and not DPI mode. In DPI mode an extra half-wave-plate is inserted into the beam path. We have at this time no evidence that this alters the astrometric solution.}.
The same fitting procedure was utilized for the NACO archival data. Since both data sets (SPHERE and NACO) were taken in H-band, we fixed the flux ratio of the two fitted Moffat functions for the NACO data set to the flux ratio extracted from the SPHERE data (see section~\ref{sec:photometry}). The astrometric calibration for NACO, taken from \cite{2010A&A...509A..52C}, gives a pixel scale of 13.24$\pm$0.05\,mas/pixel with a true north correction of -0.05$^\circ\pm$0.10. The results are listed in table~\ref{tab: astrometry} for both observing epochs. \\
We employed both astrometric epochs in order to check whether the companion candidate is co-moving with ET\,Cha on the sky. In figure~\ref{fig:pm_analysis} we show both data points relative to the expected behavior of a non-moving distant background star (grey, oscillating area in both panels). The existing astrometry is inconsistent with such an object. The dashed lines in figure~\ref{fig:pm_analysis} show the expected motion for a circular orbit. For the position angle we considered a circular face on orbit, since it would lead to the maximum change of position angle, while for the separation we considered an edge on orbit since this would lead to the maximum change in separation over time. The companion candidate shows a change in position angle larger than expected for a circular face-on orbit (see section~\ref{section:mass estimae} for a discussion of the system mass). However we also see a significant increase in separation between the two observing epochs. This likely points to an orbit with an intermediate inclination and/or a non-zero eccentricity.\\
Given that the companion candidate is inconsistent with a distant background object, we estimated the probability to find a relatively nearby, i.e. Galactic, background object within 0.15\arcsec{} of ET\,Cha and with the limiting magnitude measured for the companion candidate. Such an object could in theory exhibit a non-zero proper motion and thus could mimic a co-moving bound companion. For this we used the approach by \cite{Lillo-Box2014} and the \textit{TRILEGAL} v1.6 population synthesis models (\citealt{2012ASSP...26..165G}). We find the probability is 10$^{-6}$, i.e. negligible. Thus we conclude from the astrometric and probability analysis that the detected source is in all likelihood a true bound companion to the ET\,Cha system.

\begin{table}
 \centering
 \caption{Astrometry and photometry of the ET\,Cha system, as extracted from our SPHERE/IRDIS observations as well as NACO archival data. Note that due to the barely resolved nature of the NACO data, we did not attempt to extract the H-band photometry as it also does not add new information.}
  \begin{tabular}{@{}ccccc@{}}
  \hline 
 Epoch & Filter & Sep [mas] & PA [deg] & $\Delta$mag \\
 \hline

21-01-2003	& H & 50.5$\pm$8.3 & 45.2$\pm$6.7 & -\\
23-12-2019 & BB\_H & 135.4$\pm$0.5 & 149.7$\pm$0.8 & 1.59$\pm$0.07\\

\hline\end{tabular}
\label{tab: astrometry}
\end{table}

\subsection{Photometric analysis}
\label{sec:photometry}

We used the SPHERE/IRDIS flux calibration frames to extract relative photometry between the primary star and the companion. Since we do not have a reference PSF that is not contaminated by the close companion, we applied aperture photometry. Aperture radii of 3 pixels (36.8\,mas) were used for both objects. To achieve an accurate measurement, we estimated the cross-contamination of the companion and the primary PSF in several ways. We subtracted an azimuthally averaged profile of the primary PSF as well as a 180$^\circ$ rotated profile. We also measured the unaltered companion flux and subtracted the average background flux at the same separation but opposite side of the primary star. Between these measurements we find an $\sim$8\,\% variation in the recovered flux. We finally adopted the average value of these measurements for the companion and considered the variation in flux as uncertainty. Additionally, we included the standard deviation of the background in the uncertainty of the photometric result listed in table~\ref{tab: astrometry}. 
Since the companion is at the resolution limit in the NACO observation and was observed in the same band as the IRDIS observation, we did not attempt to extract photometry from the NACO data set.\\
To calculate the apparent magnitude of the companion we used the H-band measurement of the system listed in the 2MASS catalog (\citealt{2003yCat.2246....0C}) of 9.834$\pm$0.021\,mag. This measurement does not resolve the primary star and the companion and thus represents the combined flux. To correct for the contribution of the companion we use the formula presented in \cite{Bohn2020}. We find a correction of 0.23\,mag. Thus we compute an apparent magnitude for the companion of 11.65$\pm$0.07\,mag.

\section{The age and mass of ET\,Cha}
\label{section:age}

In order to determine the mass of the newly detected companion from photometry we need to know its age. In the following we first discuss the ET\,Cha system age, based on cluster age, kinematics and stellar parameters of the primary star. We then use this age estimate together with (sub)stellar isochrone models to determine the companion mass.

\subsection{Age estimate of the system}

The age of the $\eta$\,Cha cluster has been the subject of intense study.
In the initial study by \cite{Mamajek1999} a large spread of individual system ages was found ranging from 2\,Myr to 18\,Myr from compiled photometry, leading to an average age of $\sim$8\,Myr (\citealt{Mamajek2000}). The study by \cite{Lawson2001} broadly agrees, inferring an age range between 4\,Myr and 9\,Myr for the M-star members of the cluster from re-compiled H-R-diagrams.  
An isochronal analysis of the color-magnitude data for the cluster members by \citet{Bell2015} yielded an older cluster age of 11\,$\pm$\,3 Myr on an age scale consistent with results from Li depletion boundary analyses of other well-studied young clusters \citep[e.g.][]{Soderblom2014}.
However, we note that \cite{Herczeg2015} find a significantly younger average age of 5.5$\pm$1.3\,Myr (but with a spread between 2.1\,Myr and 12.7\,Myr) from comparison of available literature photometry and spectral types of cluster members with various stellar model isochrones.\\
The best age estimate for a cluster member is available for the RS\,Cha system (RECX 8).
Comparison of the stellar parameters for this well-constrained intermediate-mass (A8V+A8V)
eclipsing binary to modern evolutionary tracks have yielded ages of 9.1\,$\pm$\,2 Myr \citep{Alecian2007} and 8.0$^{+0.15}_{-0.25}$ Myr \citep{Gennaro2012}. \\
Due to the seemingly large age spread within the cluster it is problematic to assign the cluster age to individual sources. ET\,Cha is one of only two known members of $\eta$\,Cha with a gas rich class II disk, giving some indication that the system might in fact be younger than the average cluster age.  \cite{Woitke2011} note that the near infrared colors seen in ET\,Cha fit better a disk that is 1-2\,Myr old. 
To estimate the age of the ET\,Cha system we discuss two scenarios.\\
\\
\emph{ET\,Cha age estimate from stellar parameters:}\\
\\
Recently \cite{Rugel2018} published medium spectral resolution X-Shooter spectra of ET\,Cha taken simultaneously in the optical and near infrared. From these spectra they calculate the stellar properties and find an effective temperature of 3190\,K as well as a stellar luminosity of 0.073\,$L_\odot$. We used these values as input for \cite{Siess2000} evolutionary models as well as \cite{Baraffe2015} models. We find an age of 4.9\,Myr for the former and an even younger age of 3.2\,Myr for the latter model.  
These age estimates are on the lower end of the range for M-star cluster member proposed by \cite{Lawson2001}. To stay consistent with all age estimates we thus favor the age of 4.9\,Myr obtained from the Siess model tracks.\\
\\
\emph{ET\,Cha age estimate from kinematics:}\\
\\
Isochronal model tracks are known to underestimate the age of low mass stars (\citealt{Pecaut2012, Bell2015, Pecaut2016}). Thus ET\,Cha could be older than estimated from its stellar parameters using such models. There is in fact some compelling kinematic evidence that the system may be part of the well characterized RS\,Cha system. RS\,Cha is located only 68\arcsec{} to the North-West of ET\,Cha.
RS\,Cha shows a proper motion of -27.168$\pm$0.072\,mas/yr in RA and 28.015$\pm$0.073\,mas/yr in Dec as measured by Gaia DR2. ET\,Cha has a proper motion of -27.343$\pm$0.487\,mas/yr in RA and 27.323$\pm$0.571\,mas/yr in Dec, i.e. the motion in RA is within 1$\sigma$ of RS\,Cha and the motion in Dec is within 2$\sigma$ of RS\,Cha. Converting the differences in their proper motions vectors to tangential velocity (assuming d=100\,pc for simplicity), their tangential motions agree within 0.34$\pm$0.36 km/s. 
Using inverse Gaia parallaxes the systems are at face value located at different distances, i.e. RS\,Cha at 99.0$\pm$0.4\,pc and ET\,Cha at 91.7$\pm$2.5\,pc. However, our findings show that ET\,Cha is a close binary star with a separation of 135\,mas. Assuming a simple circular orbit for the pair we find that during the Gaia DR2 period the orbital displacement of ET\,Cha may have been of the order of 1\,mas\footnote{While the same uncertainty could affect the proper motion, here longer baselines are available that limit the influence of this deviation.}. This additional uncertainty allows for the possibility that ET\,Cha is located at slightly larger and possibly the same distance as RS\,Cha. If this is the case, then it is highly unlikely that the system is younger than RS\,Cha. 
We conclude that it currently  cannot be ruled out that ET\,Cha and RS\,Cha are forming a wide multiple system. In this case ET\,Cha should be co-eval with RS\,Cha and we adopt an age of 8\,Myr for this scenario.\\

\subsection{Mass estimates for primary and secondary}
\label{section:mass estimae}

Given the age estimate and the photometry, we can estimate the masses of the primary star and the companion in the ET\,Cha system. To compute absolute magnitudes from our photometric analysis we have to assume a distance of the system. In the case that ET\,Cha is not associated with RS\,Cha and thus 5\,Myr old, we use the Gaia DR2 parallax measurement of 91.7\,pc. However, if we assume that ET\,Cha and RS\,Cha form a wide pair then they should be located at roughly the same distance and the Gaia parallax measurement for ET\,Cha is likely flawed. For this scenario we thus adopt the distance measurement of RS\,Cha, i.e. 99.0\,pc.\\
Using BT-SETTL model isochrones for low mass stars and brown dwarfs (\citealt{Baraffe2015}), we find masses of 0.22\,M$_\odot$ and 0.048\,M$_\odot$ (50.3\,M$_{Jup}$) for primary star and companion respectively for the first scenario. Using the older age and larger distance we find values of 0.32\,M$_\odot$ and 0.10\,M$_\odot$.

\section{Limits on additional companions}

\begin{figure}
\center
\includegraphics[width=0.48\textwidth]{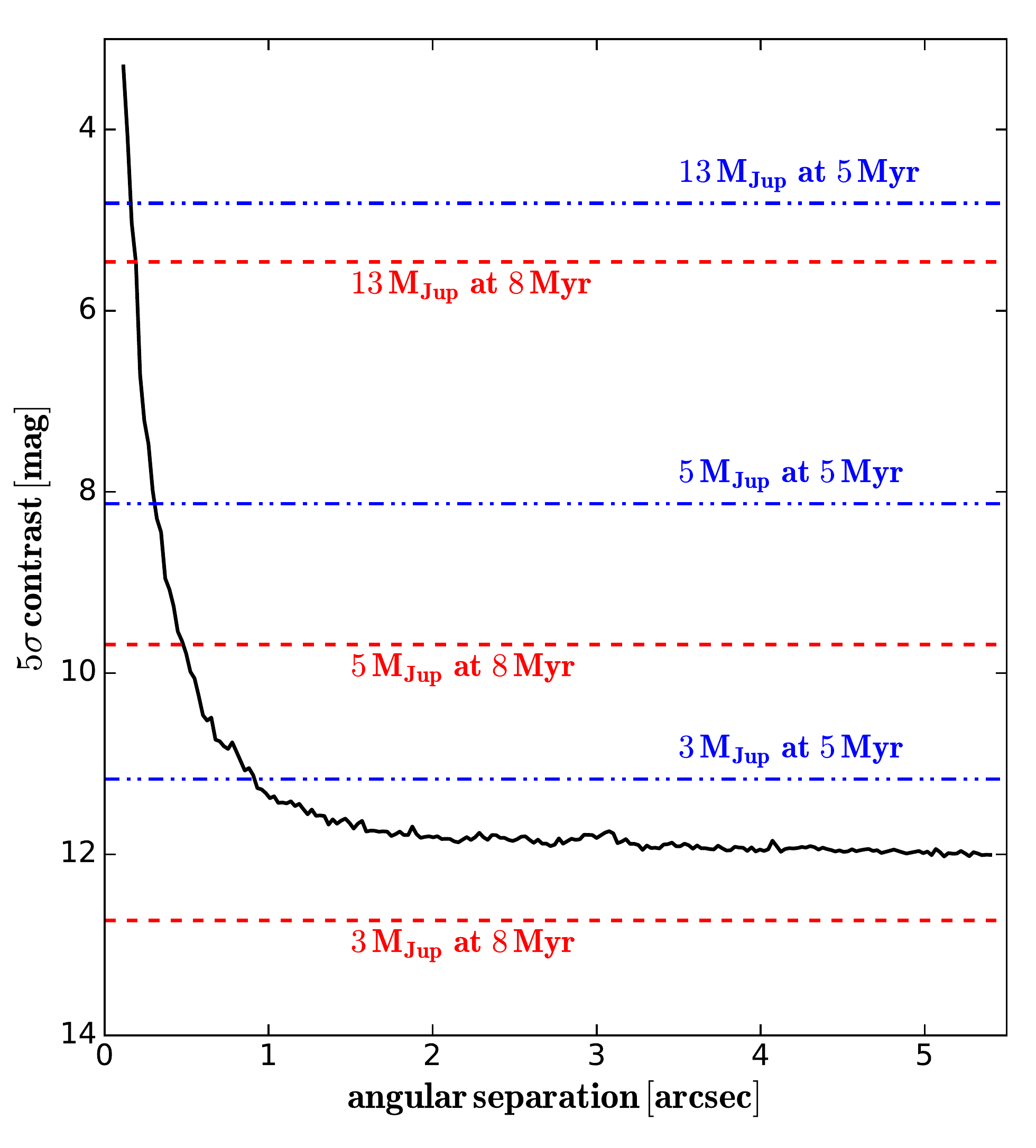} 
\caption{Contrast limits derived from the coronagraphic observations using angular differential imaging and TLOCI post-processing. Mass limits for the 5\,Myr and the 8\,Myr case are indicated with the blue, dash-dotted lines and the red, dashed lines, respectively.
} 
\label{fig:contrast}
\end{figure}

Using the deep coronagraphic total intensity images we investigated the possible presence of further companions to the ET\,Cha system.
For this purpose we applied the TLOCI angular differential imaging algorithm (\citealt{2014IAUS..299...48M}) as realised in the \textit{SpeCal} toolbox (\citealt{2018A&A...615A..92G}) implemented into the SPHERE-DC reduction pipeline (\citealt{2017sf2a.conf..347D}).
We find two additional point sources at separations of 3.03$\pm$0.02\,arcsec and 4.20$\pm$0.02\,arcsec and position angles of 82.1$^\circ \pm$0.3$^\circ$ and 135.2$^\circ \pm$0.3$^\circ$.
The closer of these is also detected in the NACO archival data and is consistent with a distant, non-moving background object (see figure~\ref{fig:pm-wide-background}). The farther source is too faint (21.3$\pm$0.2\,mag in the SPHERE H-band image) to be detected in the NACO data. We thus can not determine its nature. However, due to its wide separation it seems likely that this is a background source as well. \\
Using the procedure outlined in \cite{2018A&A...615A..92G} we use the total intensity data to determine detection limits for additional companions. The result is shown in figure~\ref{fig:contrast}. Utilizing models by \cite{Baraffe2015}, we can translate the contrast limits to mass limits. We can rule out additional stellar or brown dwarf companions down to an angular separation of $\sim$190\,mas independent of the system age. Outside of 1\arcsec{} we are sensitive to planetary mass companions down to masses of 3\,M$_{Jup}$ for the lower system age and down to 4\,M$_{Jup}$ for the higher system age.  

\section{Orbit analysis of the ET\,Cha system}
\label{sec:orbit}

We utilized the \textit{orbitize!} Python package (\citealt{2020AJ....159...89B}) to investigate possible orbit configurations of the system. We employed the OFTI (Orbits For The Impatient, \citealt{2017AJ....153..229B}) sampling method with 10$^6$ runs. 
We are considering the two scenarios for the system mass discussed in the previous section, i.e. a total mass of 0.268\,M$_\odot$ for the younger low-mass scenario 1 and a total mass of 0.42\,M$_\odot$ for the older higher-mass scenario 2. In addition we are using different distance estimates for scenario 1 and scenario 2, i.e. 91.7\,pc and 99.0\,pc respectively. Since we only have two astrometric data points and the primary goal is to get a general understanding of possible orbit families, we do not consider an uncertainty for the mass estimates in the fit, i.e. they are treated as fixed values and not free parameters. The resulting posterior distributions of semi-major axis, inclination and eccentricity are shown in figure~\ref{fig:orbit_fit}. We additionally show ten randomly selected orbits for both mass scenarios in appendix~\ref{orbit-appendix}. We note that, while we do not limit the semi-major axis to a certain parameter range, we cut off the posterior distributions shown here at 30\,au. This is motivated by \cite{Bate2009}, who find in their hydrodynamic simulations of stellar clusters that low mass binaries typically have semi-major axis smaller than 30\,au. We can however at this time not put a meaningful upper limit on the semi-major axis. Extreme eccentric solutions with very large semi-major axis up to $\sim$1000\,au are in principle consistent with the astrometric data points.\\
While we can not constrain the orbit tightly from only two data points, we find that several bound orbit families exist. We find typically either inclined or eccentric orbits or a mixture of both, and can rule out circular face-on orbits. The degeneracy between inclination and eccentricity is typical for an orbit with a low coverage of data points or with only short orbital arcs observed (see e.g. \citealt{2014MNRAS.444.2280G}).\\ 
\\
\emph{Low-mass scenario 1:}\\
\\
For the low-mass scenario 1 we find a first orbit family with the most likely inclination range between 30$^\circ$ and 73$^\circ$, i.e. the inclination is rather unconstrained.
These solutions can be circular, but have the strongest probability peak between eccentricities of 0.15 and 0.3. The most likely semi-major axis range is 9\,au to 15\,au with a peak at 13\,au.
A second orbit family favors high eccentricity values of roughly 0.2 to 0.75 which correspond to slightly smaller semi-major axes between 8\,au and 11\,au with peak at 8.5\,au. These solutions have a smaller inclination roughly between 0$^\circ$ and 40$^\circ$.
We find a general lower limit of the semi-major axis across all solutions of 6.5\,au, but can neither constrain inclination nor eccentricity to any upper or lower values. \\
\\
\emph{High-mass scenario 2:}\\
\\
For the higher-mass scenario 2 we find similarly two orbit families. The high inclination family shows a probability peak in the inclination between 55$^\circ$ and 75$^\circ$. These solutions have a most likely semi-major axis range between 10\,au and 19\,au with a strong peak at 14\,au. These solutions can be circular and have the strongest probability between eccentricities of 0 and 0.2. Compared to the lower mass scenario we thus find that for this first orbit family high inclinations, larger semi-major axis, but lower eccentricities are preferred. 
The second orbit family are the more eccentric solutions with a probability peak in the eccentricity space between 0.4 and 0.7. These solutions have smaller semi-major axes with a peak between 9\,au and 10\,au. As was the case for the lower-mass scenario these solutions have smaller inclinations roughly between 0$^\circ$ and 40$^\circ$. Thus this second orbit family is located in a very similar parameter space to the lower-mass scenario.\\
While this first assessment of the system orbit is instructive, we caution that this picture might change significantly with the addition of even one well calibrated observing epoch.\\

\begin{figure}
\center
\includegraphics[width=0.48\textwidth]{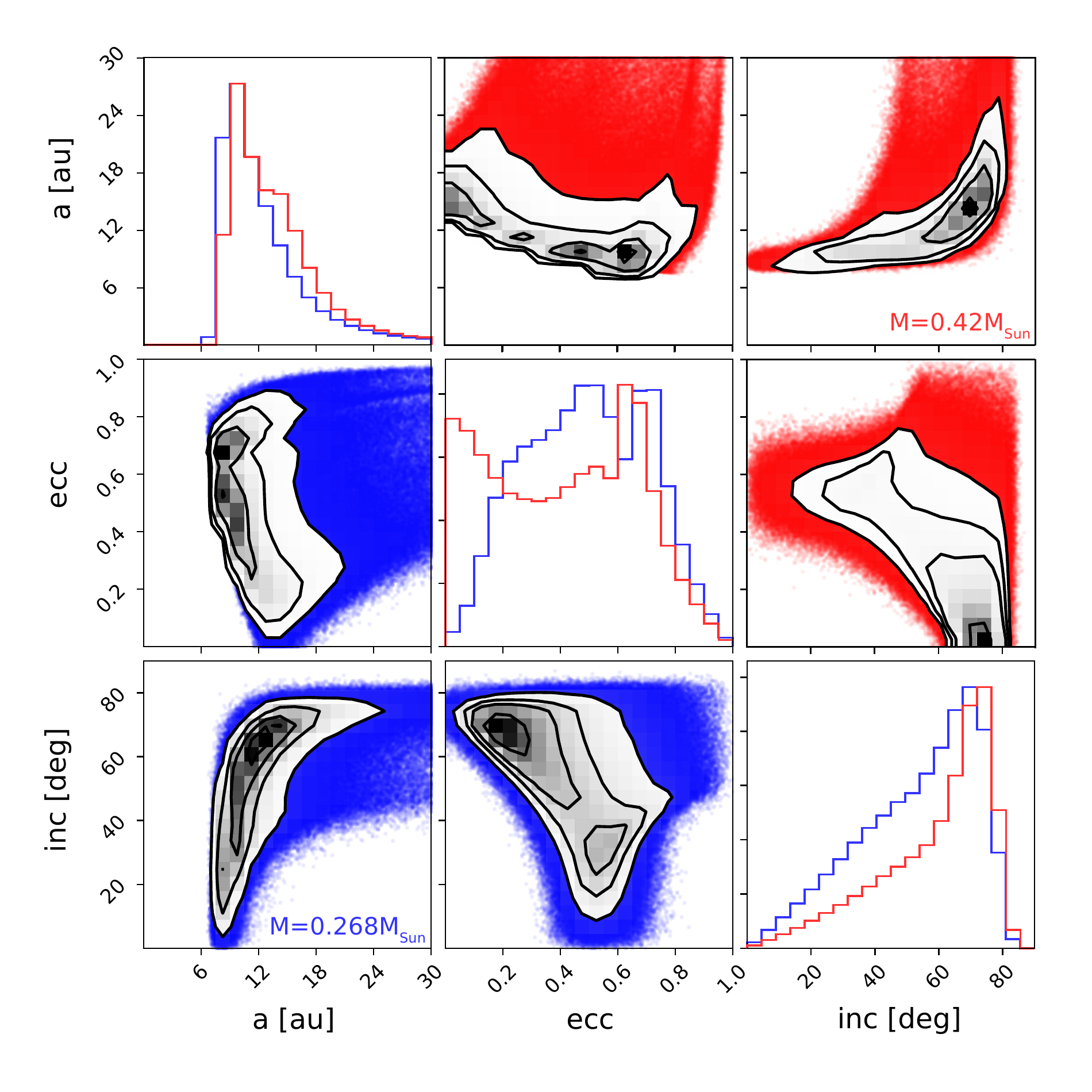} 
\caption{Resulting orbit solutions for the ET Cha system using our extracted astrometry.
We utilized the orbitize! package and the included OFTI algorithm with 10$^6$ generated orbits.
We show semi-major axis, inclination and eccentricity.
Scenario 1 is in the lower left corner in blue color and scenario 2 is in the upper right corner in red color.
} 
\label{fig:orbit_fit}
\end{figure}

\section{Discussion}

Our photometric and age analysis finds that the companion is either a low-mass (0.10\,M$_\odot$) pre-main-sequence M-type star or a brown dwarf (0.048\,M$_\odot$, i.e. 50.3\,M$_{Jup}$) depending on the system age. Accordingly, the mass ratio between primary star and companion is either 0.31 or 0.21 (with the primary star itself also having an age dependent mass). In both cases this makes for a somewhat atypical system. \cite{Bate2009} found with hydrodynamic simulations of stellar clusters that the median mass ratio for binary systems with a semi-major axis smaller than 10\,au is 0.74 and for systems with semi-major axis between 10\,au and 100\,au is 0.57. This theoretical result is supported also by observational surveys, e.g. \cite{Delfosse2004} find that brown dwarf companions are rare within 100\,au from main-sequence M-dwarf primary stars ($\sim$1\,\% of their sample stars had a brown dwarf companion). See also \cite{2013ARA&A..51..269D} and references therein, where similar results are discussed for pre-main-sequence stars. On the other hand \cite{Bate2009} also finds that the separation between binary components depends strongly on the primary mass, i.e. it increases with increasing mass. For primary masses between 0.2\,M$_\odot$ and 0.5\,M$_\odot$ they find a bi-modal distribution with roughly half of the recovered systems exhibiting semi-major axis smaller than 10\,au, i.e. compatible with a large fraction of the recovered orbit solutions for ET\,Cha.\\

\subsection{The formation of the ET\,Cha system}

There are several possible formation pathways for systems like ET\,Cha. The most prominent ones are either fragmentation in the proto-stellar cloud (e.g., \citealt{Bate1995,Kroupa1995,Lomax2015,Moe2019}) and gravitational instability in the proto-stellar or circumstellar disk (e.g., \citealt{1997Sci...276.1836B,Kratter2016}). Both of these mechanisms will (at least initially) produce dynamically very different systems. While an object formed via cloud fragmentation can show strong spin-orbit misalignment and potential high orbit eccentricities, this would not be expected from an object formed in a disk around the primary star. Both mass scenarios for the ET\,Cha system produce an appreciable number of eccentric orbit solutions. These are in particular preferred for the lower mass scenario in which the system is younger and the new detected companion in the brown dwarf regime.\\ 
However, objects formed via fragmentation in a disk may also exhibit eccentric orbits if they experienced dynamic encounters. \cite{Reipurth2001} suggest scattering of low mass cores in multiple systems as a main formation pathway to explain wide orbit brown dwarfs. For this a third body would be needed, which is so far not observed in the ET\,Cha system. Such a body could either be in a close orbit around ET\,Cha, it could have fallen into ET\,Cha\,A or it could potentially be another cluster member. The possibility that ET\,Cha is bound to RS\,Cha would make this a complex multiple system. A dynamical scattering of the ET\,Cha system by another cluster member may also be supported by findings of \cite{Moraux2007}. They simulated the $\eta$\,Cha cluster and found that ejection of cluster members may have occurred with most objects ejected in the early stages of formation, after roughly 1-4\,Myr. In their simulation they find ejection velocities of 1-5\,$kms^{-1}$, which translates into a distance of 9\,pc from the cluster core after 7\,Myr. Such ejected members of the $\eta$\,Cha cluster were indeed found by \cite{Murphy2010}. They suggest a halo of low mass members of the cluster within 5$^\circ$.5 from the cluster center, i.e. within 9\,pc. In this scenario ET\,Cha would be a member of this low-mass halo which was ejected towards us. If the Gaia parallax is taken at face value it may support such a history of ET\,Cha since it is located roughly 7\,pc closer than the median of the $\eta$\,Cha cluster (see figure~\ref{fig:parallax}). However, we caution that such a scenario in which the system is ejected directly toward us seems unlikely. \\
Besides the dynamical signatures there are several numerical studies that give some evidence to the formation history of the system. Most notably \cite{Vorobyov2013} performed simulations of disk gravitational fragmentation and found that they were unable to produce brown dwarf companions at small orbital separations. Furthermore \cite{Kratter2010}, \cite{Offner2010} and \cite{Haworth2020} found that disk fragmentation is less likely around M-dwarf primary stars. We thus suggest that there exists some circumstantial evidence that ET\,Cha\,B indeed formed via core fragmentation in the proto-stellar cloud. 

\subsection{Interaction with the circumstellar disk}

The circumstellar disk around the primary star was not detected in our scattered light observations down to an average\footnote{As noted previously the mask was misaligned, thus we probe closer to the star in the North-West and slightly further away from the star in the South-East.} separation of 0.0925\arcsec{} (8.5-9.2\,au, depending on the system distance), i.e. the nominal inner working angle of the employed coronagraph. This confirms its previously inferred small radius (5-7\,au, see \citealt{Woitke2019}). ALMA surveys in the past years have shown that such small disks are not uncommon (see e.g. \citealt{Ansdell2016}). Given the newly detected close B component the small size of the disk is indeed not surprising and is likely explained by truncation. In such a case the expected disk outer radius is half the periastron separation (\citealt{Hall1996}). The closest projected separation was observed with NACO to be 50.5\,mas. If we assume that this is the actual physical separation at periastron, then the disk should have been truncated at 2.3-2.5\,au. However, this assumes that the entire orbital trajectory is in the plane of the sky, which might well not be the case (we recover many orbits with larger periastron separations). So this should be seen as a lower limit and is in principle consistent with the inferred disk radius of 5-7\,au.
Truncation by outer companions is indeed common. \cite{Manara2019a} found recently that disks in  known multiple systems are systematically smaller in mm continuum emission than their counter parts around single stars. \\
The disk around ET\,Cha is still unusual in several aspects. \cite{Kraus2012} find from an observational study in Taurus that the disk frequency is significantly reduced around close ($\leq$50\,au) binaries. While they find disks in more than $\sim$80\,\% of wide binaries (same result as for single stars), this is true for only $\sim$40\,\% of close binaries. These results are supported by recent population synthesis models by \cite{Rosotti2018}, who find that binaries with separations similar to ET\,Cha ($\sim10\,au$) only have a disk in 10\,\% of the cases. In the same study they predict that in these close systems the disk around the secondary component will clear first, in line with our non-detection of a resolved disk around the B component.\\
A second puzzling aspect of the system is its high accretion rate. Using the UV excess measurement \cite{Rugel2018} estimated an accretion rate of 7.6$\times10^{-10}$M$_\odot yr^{-1}$. Assuming a gas mass of 1.2$\times10^{-4}$\,\,M$_\odot$ (\citealt{Woitke2019}) and a constant accretion rate the circumstellar disk should be gone after only $\sim$ 1.6$\times10^{5} yr$, i.e. a time much shorter than both our estimates for the system age.
However \cite{Rosotti2018} found that a close companion has significant influence on the evolution of the disk. In particular for a semi-major axis smaller than 20-30\,au the dominating disk dispersal mechanism changes from the inside-out regime (through photo-evaporation) to the outside-in regime due to the tidal torque of the companion. Thus in these disks no inner cavity is opened, which leads to significantly higher accretion rates that for wide separation binary stars or single stars. In particular the dimensionless $\eta$ parameter that was studied by \cite{Rosotti2017} (see also \citealt{Jones2012}) and that is the product of system age and accretion time divided by disk mass, shows a steep increase with age for these systems. For ET\,Cha we compute values for $\eta$ of 31 and 51 for the younger and older disk age. Such values are possible in the simulations by \cite{Rosotti2017}, but given the system separation they imply an age younger than 1.5\,Myr for a value of the viscous parameter $\alpha$ (\citealt{Shakura1976}) of 10$^{-3}$. \\
To reconcile the age and accretion rate of ET\,Cha we require a lower viscous parameter $\alpha$ of 10$^{-4}$. This would increase the disk viscous timescale at a truncation radius of 10\,au to 5\,Myr. Since the disk dispersal takes on the order of 2-3 viscous timescales (\citealt{Pringle1981, Rosotti2018}) the presence of the disk in both age scenarios would then not be problematic. However lowering the viscous timescale would also imply that we need a higher disk mass to explain the current high accretion rate (assuming purely viscous accretion). We roughly find that an increase by a factor 10-15 would be required. The disk mass could be indeed significantly higher than inferred by \cite{Woitke2019} if the disk is optically thick outside of 1\,au. \\
We note that very recently \cite{Manara2020} found similarly high accretion rates as reported for ET\,Cha around several members of the $\sim$5\,Myr Upper Scorpius region (see also \citealt{Ingleby2014,Venuti2019} for Orion OB1 and TWA). The mass estimate in this case was based on the dust. They suggest that a higher gas-to-dust ratio than the often assumed 100 would explain the measured accretion rates. Indeed \cite{Woitke2019} find with their thermo-chemical modeling of the ET\,Cha system an extreme gas-to-dust ratio of 3500 and the true value would be even more extreme if the gas mass is indeed underestimated. However it would be very interesting to study the sample of \cite{Manara2020} with high angular resolution to test the correlation of a high accretion rate with the occurrence rate of close companions. \\
We find that another scenario might simultaneously explain the discrepancy of the age and accretion rate of the system as well as the small size of the circumstellar disk. If the companion is not bound but instead is on a hyperbolic orbit, i.e. we are imaging the system close to the periastron passage during a fly-by, then the disk could have been recently truncated. However, we do not see evidence for a dispersing disk outside of the companion orbit. Also such a scenario is inherently unlikely because close encounters are rare (\citealt{2006ApJ...641..504A,2018MNRAS.478.2700W}) and it is even more unlikely to observe them close to periastron passage. We nevertheless include for completeness that with only two astrometric epochs we can not rule out a hyperbolic orbit (even though we did not specifically fit unbound trajectories).  \\
Finally it may be possible that the accretion rate is highly variable if accretion "pulses" are triggered by the companion during periastron passage of an eccentric orbital trajectory (e.g., \citealt{Tofflemire2019}). 

\section{Summary and conclusion}

We detected a low mass (50.3\,M$_{Jup}$ or 0.1\,M$_\odot$, depending on system age) companion to the $\eta$\,Cha cluster member ET\,Cha.
This companion is inconsistent with a background object and in all likelihood associated with ET\,Cha. From SPHERE and NACO measurements spaced almost 17 years apart, we can see significant orbital motion, which can be explained by several families of bound orbits, many of them with significant eccentricity. Due to a lack of additional data points we can however not rule out hyperbolic orbits.\\
The mass ratio of the system is low compared to theoretical and observational studies, possibly representing an extreme case of a young multi star system.
From the small separation, low mass ratio and potential eccentric orbit we tentatively conclude that the companion may have formed via fragmentation in the proto-stellar cloud.\\
The disk around ET\,Cha has several characteristics, such as its small outer radius, its high gas-to-dust ratio and a high accretion rate compared to age and gas mass, which may all well be explained by the companion. In particular the small separation of the pair indicates that the disk clearing might be dominated by tidal torques from the companion, which also trigger the high accretion rate. If we assume purely viscous accretion than we find that we need a low $\alpha$ of $\sim$10$^{-4}$ to explain the presence of he disk at the age of the system. This is in line with with recent studies of multi-ringed disks, which require also a low viscosity (\citealt{Dullemond2018ApJ}).
To come to more definite conclusions regarding the evolutionary state of the system and its dynamical history follow-up observations are required. In particular we suggest the following:

\begin{enumerate}
    \item SPHERE/IRDIS follow-up observations spread over the next few years to determine the orbit of the system. In particular if the orbit is bound and if so if it is highly eccentric.
    \item Search for accretion tracers of the companion, which may indicate in-situ formation or very recent ejection from the inner system. This may be done with SPHERE/ZIMPOL or VLT/MUSE in H$\alpha$ or possibly MagAO-X once it is online.
    \item Non-coronagraphic follow-up observations with SPHERE/IRDIS to determine the polarization state of both objects and thus infer (or rule out) the presence of circumstellar material around the companion. 
    \item VLT/ERIS measurements (once it is available) to get a companion spectrum and possibly its radial velocity, which would significantly constrain its orbit as well as mass.
    \item Very high spatial resolution (about 3\,au should be possible) and sensitive ALMA observations of the gas and dust. Such observations can provide stringent  constraints on the gas and dust mass,  the extension of the disk, and on the presence of a (remnant) circumbinary disk. Furthermore the spectral line observations, with the necessary spectral resolution, can provide additional constrains on the mass of the primary.  

\end{enumerate}

Stellar multiplicity in general can have a strong influence on the evolution of circumstellar disks. Our new observations show that with extreme adaptive optics instruments it is now possible to detect previously unnoticed (sub)stellar companions to young stars, in particular in a parameter range (separation and mass) where they may cause significant changes in disk evolution. Currently instruments such as SPHERE and GPI, are limited in their target sample by the requirements of optical bright guide stars for their adaptive optics systems. This leads to an observational bias towards the higher end of the mass function. This is one of the many reasons why instrument upgrades such as the proposed SPHERE+ concept (\citealt{2020arXiv200305714B}) are highly important to understand the evolution of young systems.


\section*{Acknowledgements}

The authors would like to thank Anthony Brown for fruitful discussions.
SPHERE is an instrument designed and built by a consortium
consisting of IPAG (Grenoble, France), MPIA (Heidelberg, Germany),
LAM (Marseille, France), LESIA (Paris, France), Laboratoire Lagrange
(Nice, France), INAF - Osservatorio di Padova (Italy), Observatoire de
Genève (Switzerland), ETH Zurich (Switzerland), NOVA (Netherlands), ONERA
(France), and ASTRON (The Netherlands) in collaboration with ESO.
SPHERE was funded by ESO, with additional contributions from CNRS
(France), MPIA (Germany), INAF (Italy), FINES (Switzerland), and NOVA
(The Netherlands). SPHERE also received funding from the European Commission
Sixth and Seventh Framework Programmes as part of the Optical Infrared
Coordination Network for Astronomy (OPTICON) under grant number RII3-Ct2004-001566
for FP6 (2004-2008), grant number 226604 for FP7 (2009-2012),
and grant number 312430 for FP7 (2013-2016). 
C.G. acknowledges funding from the Netherlands Organisation for Scientific Research (NWO) TOP-1 grant as part
of the research program “Herbig Ae/Be stars, Rosetta stones for understanding
the formation of planetary systems”, project number 614.001.552.
EEM acknowledges support from NASA award 17-K2GO6-0030.
Part of this research was carried out at the Jet Propulsion Laboratory, California Institute of Technology, under a contract with the National Aeronautics and Space Administration (NASA). 
FMe acknowledges funding from ANR of France under contract number ANR-16-CE31-0013.
GR acknowledges funding from the Dutch Research Council (NWO) with project number 016.Veni.192.233.
R. A.-T. acknowledges support from the European Research Council under the 
Horizon 2020 Framework Program via the ERC Advanced Grant Origins 83 24 28.
T.B. acknowledges funding from the European Research Council under the European Union’s Horizon 2020 research and innovation programme under grant agreement No 714769 and funding from the Deutsche Forschungsgemeinschaft under Ref. no. FOR 2634/1 and under Germanys Excellence Strategy (EXC-2094–390783311).
This work has made use of the the SPHERE Data Centre, jointly operated by OSUG/IPAG (Grenoble),
   PYTHEAS/LAM/CESAM (Marseille), OCA/Lagrange (Nice), Observatoire de Paris/LESIA (Paris),
   and Observatoire de Lyon. 
This research has used the SIMBAD database, operated at CDS, Strasbourg, France \citep{Wenger2000}. 
We used the \emph{Python} programming language\footnote{Python Software Foundation, \url{https://www.python.org/}}, especially the \emph{SciPy} \citep{2020SciPy-NMeth}, \emph{NumPy} \citep{oliphant2006guide}, \emph{Matplotlib} \citep{Matplotlib}, \emph{photutils} \citep{photutils}, and \emph{astropy} \citep{astropy_1,astropy_2} packages.
We thank the writers of these software packages for making their work available to the astronomical community.

\bibliographystyle{aa}
\bibliography{MyBibFMe}

\newpage
\begin{appendix}

\section{Proper motion test for wide separation companion}

  \begin{figure}[h!]
  \resizebox{\hsize}{!}{\includegraphics[]{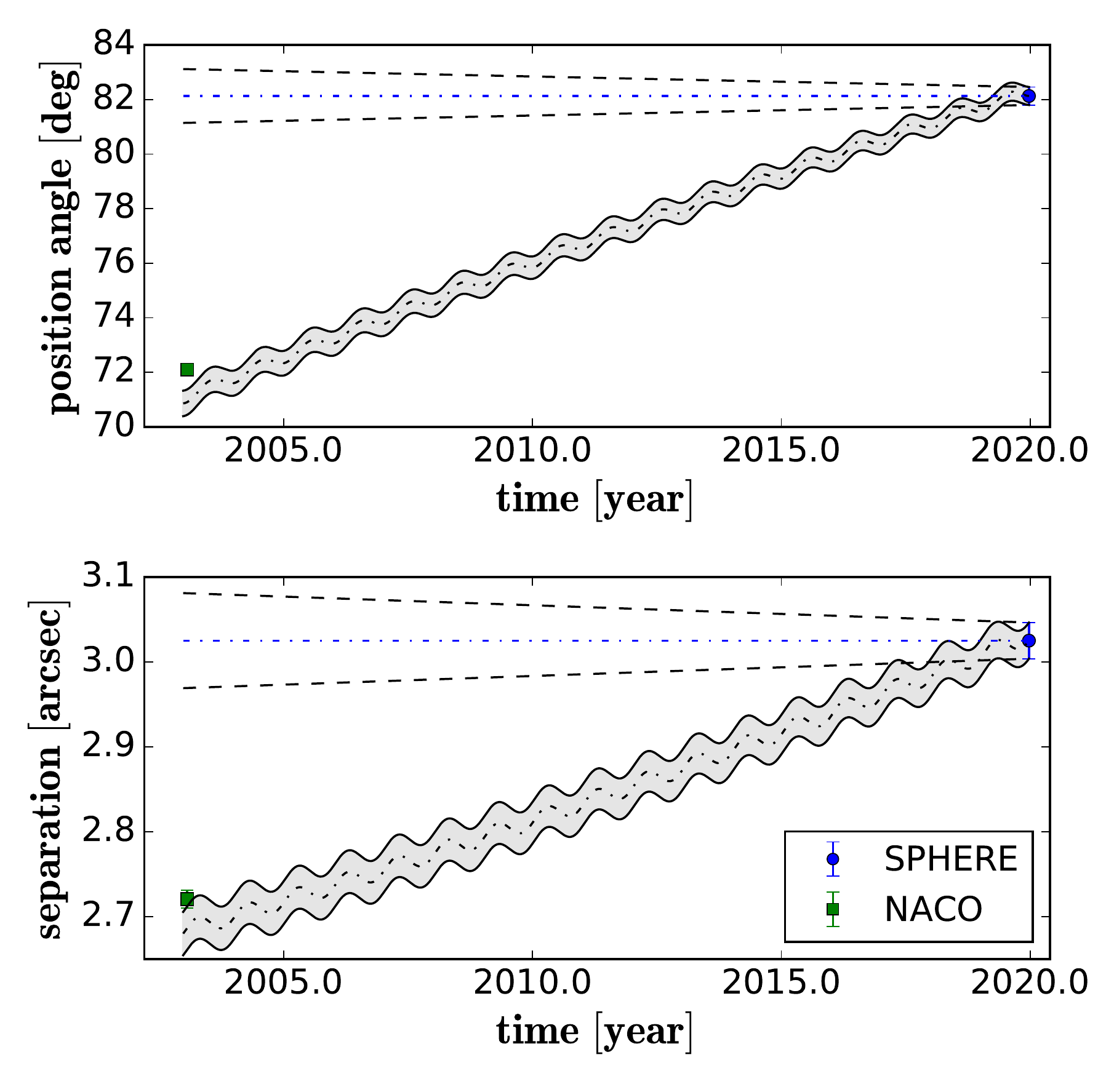}}
      \caption{Astrometric proper motion analysis analogues to figure~\ref{fig:pm_analysis}, but for the additional wide separation point source detected in the coronagraphic images. The astrometry in the SPHERE and NACO epochs is consistent with a non-moving (distant) background object.}
         \label{fig:pm-wide-background}
  \end{figure}

\newpage

\section{Randomly selected orbit plots}
\label{orbit-appendix}
To illustrate the quality of the recovered orbit solutions in section~\ref{sec:orbit} we show for both mass scenarios 10 randomly selected orbits. Astrometric data points are displayed in black. 

\begin{figure*}[h!]
\center
\includegraphics[width=0.98\textwidth]{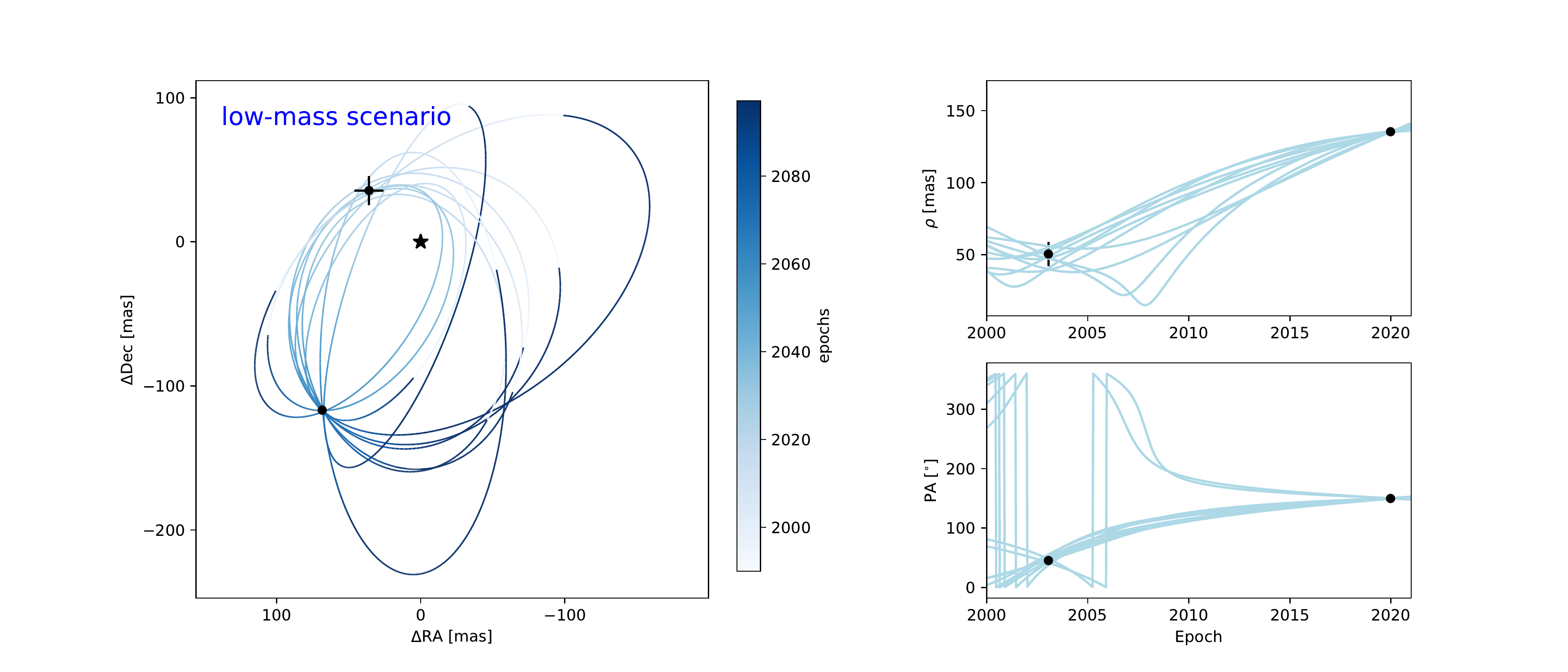} 
\caption{Ten random orbits for the low-mass scenario 1, i.e. a system mass of 0.268\,M$_\odot$. On the left we show the orbit in RA-Dec space and on the right we show relative separation and position angle of the secondary relative to the primary versus time.
} 
\label{fig:app:coro}
\end{figure*}

\begin{figure*}[h!]
\center
\includegraphics[width=0.98\textwidth]{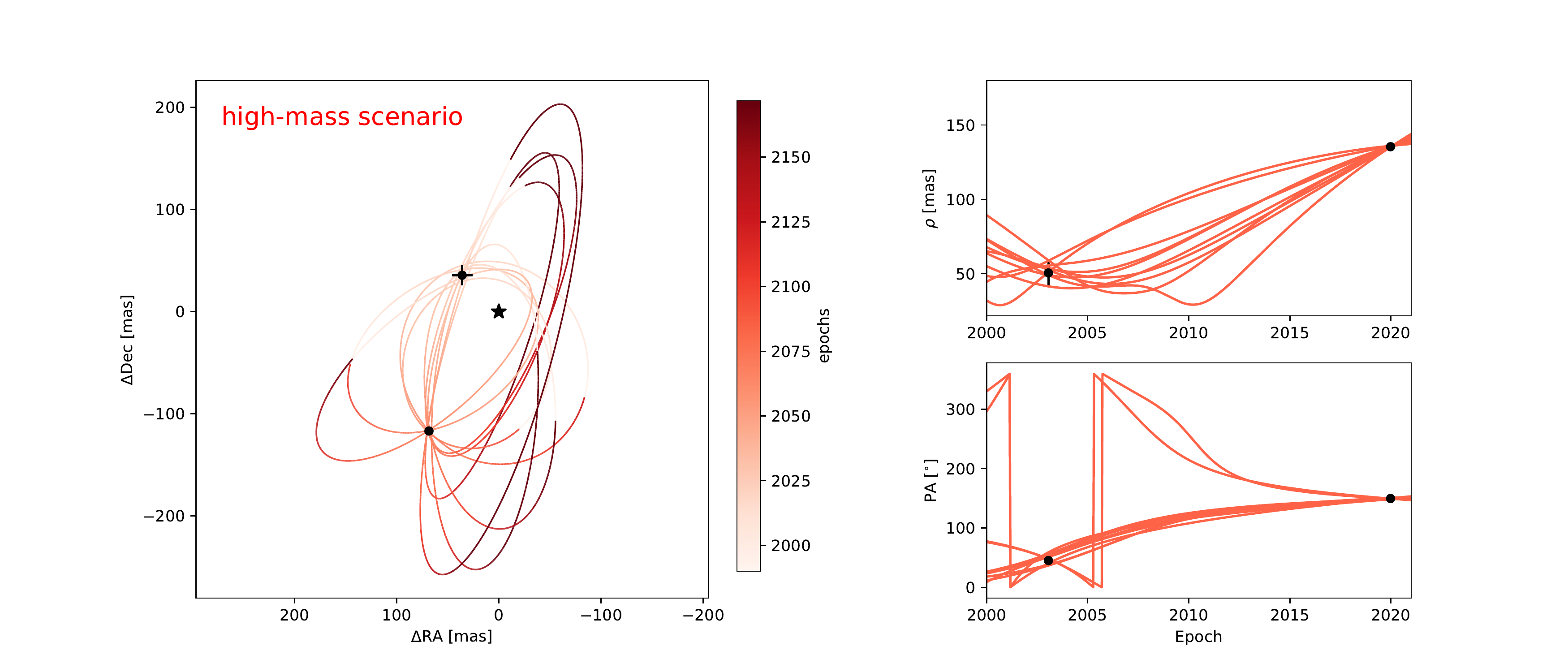} 
\caption{Ten random orbits for the high-mass scenario 2, i.e. a system mass of 0.42\,M$_\odot$. On the left we show the orbit in RA-Dec space and on the right we show relative separation and position angle of the secondary relative to the primary versus time.
} 
\label{fig:app:coro}
\end{figure*}

\newpage

\section{Cluster parallaxes}

  \begin{figure}[h!]
  \resizebox{\hsize}{!}{\includegraphics[]{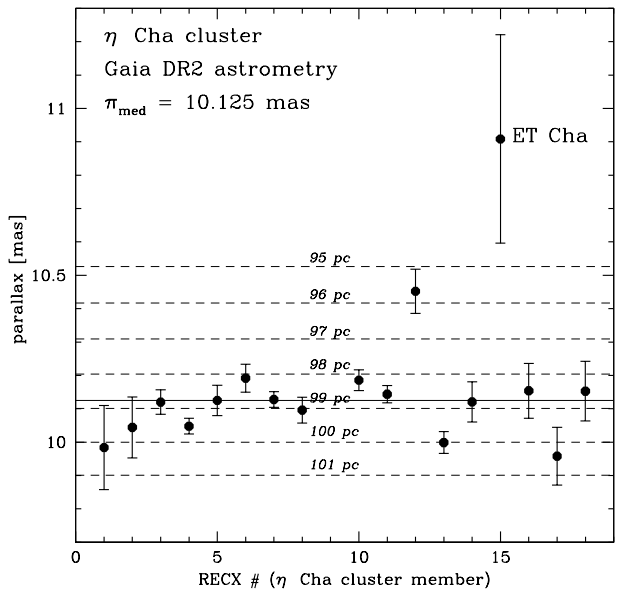}}
      \caption{Parallaxes of known $\eta$ Cha cluster members. ET\,Cha seems to be significantly closer than the other members.}
         \label{fig:parallax}
  \end{figure}

\end{appendix}

\end{document}